\documentclass[alpha-refs]{wiley-article-arxiv}

\usepackage{graphicx}
\usepackage[space]{grffile}
\usepackage{latexsym}
\usepackage{textcomp}
\usepackage{longtable}
\usepackage{tabulary}
\usepackage{booktabs,array,multirow}
\usepackage{amsfonts,amsmath,amssymb}
\usepackage{natbib}
\usepackage{url}
\usepackage{hyperref}
\hypersetup{colorlinks=false,pdfborder={0 0 0}}
\usepackage{etoolbox}

\newcommand{\cd}{\,|\,}
\newcommand{\ci}{\mbox{\protect $\: \perp \hspace{-2.3ex}
\perp$ }}

\newcommand{\nn}[0]{\hspace*{.7em}}

\newcommand{\arc}{\mbox{$\hspace{.06em} \prec
\!\!\!\!\!\frac{\nn \nn}{\nn}
\!\!\!\!\!
\succ\! \hspace{.25ex}$}}

\newcommand{\cov}{\mathrm{cov}}
\newcommand{\hub}{\mathrm{hub}}

\makeatother
\newif\iflatexml\latexmlfalse

\AtBeginDocument{\DeclareGraphicsExtensions{.pdf,.PDF,.eps,.EPS,.png,.PNG,.tif,.TIF,.jpg,.JPG,.jpeg,.JPEG}}

\usepackage[utf8]{inputenc}
\usepackage[english]{babel}

\usepackage{siunitx}

\iflatexml

\else

\corraddress{Department of Statistical Science, University College, Gower Street London, WC1E 6BT, United kingdom}
\corremail{k.sadeghi@ucl.ac.uk}
\fundinginfo{AFOSR, Grant Number: FA9550-14-1-014}
\fi

\title{Hierarchical Models for Independence Structures of Networks}

\author[1]{Kayvan Sadeghi}

\affil[1]{University College London}

\author[2]{Alessandro Rinaldo}

\affil[2]{Carnegie Mellon University}

\runningauthor{Sadeghi and Rinaldo}

\begin{document}

\maketitle
\selectlanguage{english}
\begin{abstract}
We introduce a new family of network models, called
hierarchical network models, that allow us to represent in an explicit manner
the
stochastic dependence among the dyads (random ties) of the network. In particular, each
member of this family can be associated with a graphical model defining conditional independence clauses among the dyads of the network, called the dependency graph. Every network
model with dyadic independence assumption can be generalized to construct
members of this new family. Using this new framework, we generalize the
Erd\"{o}s-R\'{e}nyi and $\beta$-models to create hierarchical
Erd\"{o}s-R\'{e}nyi and $\beta$-models. We describe various methods for parameter estimation as well as simulation studies for models with sparse dependency graphs. 

\end{abstract}%

\section{Introduction}
The statistical analysis of network data is concerned with modeling relational
data taking the form of random graphs, or networks, where the nodes represent
units in a population of interest and the random ties (i.e.\ dyads) encode the complex of
interactions among them.

Many well-known network models and, in fact, most models that can be fit to large networks
either rely  on the often unrealistic but theoretically convenient assumption of \emph{dyadic independence}, that is probabilistic independence among the dyads, or induce complete dependence among the  dyads.
Developing network models that explicitly accounts for more complex forms of marginal or conditional independencies among the dyads, also called the \emph{independence structure of the network}, has proved to be quite difficult, both for computational and theoretical reasons.  As a result,  most of the network models proposed in the literature and used in practice allow only for a minimal degree of control over the type and strength of stochastic dependence among dyads. 

On the other hand, in the field of graphical models, there is a vast body of work
on the independence structure of sets of random variables. In graphical models,
graphs, generally called \emph{Markov dependency graphs} or simply \emph{dependency graphs}
 consist of vertices that correspond to random variables and edges that
correspond to some types of conditional dependencies between their endpoints;
see for example \citet{lau96}. Throughout, in order to avoid confusion, we use the terms ``node" and ``tie" for networks, and ``vertex" and ``edge" for dependency graphs, whereas a potential random tie is called a ``dyad". In order to distinguish between different
possible conditional independencies among random variables, different typologies
of dependency graphs have been proposed in the graphical model literature, each of which
corresponding to a particular class of models. One of the most popular class of
graphical models is that of \emph{undirected graph models} \citep{dar80}, which
assume conditional independence between random variables corresponding to two
non-adjacent vertices in the dependency graph given all other variables (vertices of the graph).

There is a natural duality between networks and dependency graphs: dyads of the network are (binary) random variables, and hence can be considered to be  vertices of a dependency graph. Edges of the
dependency graph would then determine the conditional independence among these variables, i.e.\ the independence structure among the dyads of the network.
This duality was first noted and used for modeling purposes in the seminal work of
\citet{fra86}, where network dyads are modeled using a very specific type of dependency graph that assumes dyads to be conditionally independent if they do not share a node; see also Chapter 7 of \citet{lus12}. Recently, this duality was used for sampling and model fitting in certain network models in \citet{thi17}, or for modeling \emph{exchangeable} random networks in \citet{lau17,lau19}. Also, there have been other approaches to deal with certain ``local" types of dependency in networks by using \emph{nodal attributes}, i.e.\ extra information on the individuals presented by nodes of the network \cite{sch12} and  \citet{fel12}.

The main goal of this paper is to leverage the duality between networks and
dependency graphs in order to propose new, tractable and interpretable  network models that allow for specific independence structures.  We accomplish this by generalizing existing and well-known network models that rely on the dyadic independence assumption to hierarchical log-linear models that additional conform to a given set of conditional independencies among the dyads.
Towards that end, we work
with undirected graphical models, which imply that our proposed network models
are of linear exponential family form and, therefore, are instances of the class of \emph{exponential random graph models} (ERGMs); see \citet{hol81}
\citep[but see also its discussion,][]{fie81} and \citet{fra86}, and, for recent developments, \citet{hun06,rob07}.
As baseline models, we  consider both the
Erd\"{o}s-R\'{e}nyi models, defined by \citet{erd59} and studied vastly in the
literature of networks and random graph theory and the $\beta$-models, defined by \citet{bli10,cha11} and studied by \citet{rin13}. We call the resulting models the hierarchical Erd\"{o}s-R\'{e}nyi and hierarhical $\beta$ model respectively. Our approach could also accommodate, in the same manner, for the class of $p_1$ models, introduced by \citet{hol81} for directed networks.

We also provide a method based on the gradient descent algorithm to estimate the  maximum of the likelihood function for hierarchical Erd\"{o}s-R\'{e}nyi models. In principle, this method can be generalized to other families of hierarchical models, but the computational difficulties should be examined in more detail. We also provide simulation studies to apply the proposed method for the maximum likelihood estimation, and to compare hierarchical  Erd\"{o}s-R\'{e}nyi with Erd\"{o}s-R\'{e}nyi.


In the next section, we formally define networks and dependency graphs, introduce the independence structures for undirected  dependency graphs, and illustrate the duality between networks and dependency graphs.

In Section 3.1, we introduce hierarchical log-linear models for undirected
graphical models, and demonstrate how to use them to define hierarchical
network models.
In Section 3.2,  we apply the Erd\"{o}s-R\'{e}nyi
parametrization to hierarchical network models, defined in Section 3.1, write the
models in exponential family form, and provide the corresponding normalizing
constant in closed form in the sense that they do not depend on summation over all
networks. In Section 3.3, we conduct a similar study as in
section 3.2 for the $\beta$-model parametrization instead of Erd\"{o}s-R\'{e}nyi's.  In Section 4, we study the maximum likelihood estimation and provide algorithms for this purpose for (sparse) dependency graphs for hierarchical  Erd\"{o}s-R\'{e}nyi. We also provide relevant simulation studies, which show that the models significantly take the simulated dependencies in the networks into account.

In Section 5, we present several problems related to the proposed models for further work. In particular we briefly discuss  model selection for the dependency graph, and the existence of the maximum likelihood estimator for these models, as well as the selection of dependency graphs. We finally discuss analogous models based on marginal independence. 
\section{Networks and dependency graphs}\label{sec:gr.mod}
Graphical models \citep[see, e.g.][]{lau96} are statistical models expressing
conditional independence statements among a collection of random variables $X=(X_1,\dots,X_{|N|})$ indexed by a finite set $N$. A graphical model is determined by a \emph{dependency graph} $D=(N,F)$ over the
indexing set $N$, whose edge set $F$ (which may include edges of undirected, directed
or bidirected type) encodes conditional independence
relations among the variables, or {\it Markov properties}. For a non-empty set $A \subset N$, let $X_A = (X_i, in \in A)$ be  the corresponding  sub-vector of $X$. Given disjoint subsets  $A$, $B$ and $C$ of $N$, with $A$ and $B$ non-empty, we express the clause that $X_A$ is conditional independent of $X_B$ given $X_C$ using the notation $A\ci B\cd C$. In particular, if $C$ is the empty set, this will reduce to simply marginal independence of $X_A$ and $X_B$, written as
$A\ci B$.

For an \emph{undirected graph} $D$, where all edges are depicted as full lines, if, for any two non-adjacent vertices $i$ and $j$, it holds that $i\ci j\cd V\setminus\{i,j\}$, i.e.\ $i$ and $j$ are conditionally independent given the rest of the vertices then we say that the \emph{pairwise Markov property} is satisfied. If, for any three disjoint subsets $A$, $B$ and $C$ of the vertex set, $A\ci B\cd C$ when every path between $A$ and $B$ has a vertex in $C$ then we say that the \emph{global Markov property} is satisfied. It is known that these two conditions are equivalent for positive densities; see \citet{pea88,lau96}.

\begin{figure}[h!]
             \centering
             \scalebox{0.55}{\includegraphics{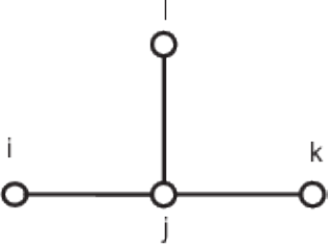}} 
             \caption{An undirected dependency graph.}
         \label{fig:001}
 \end{figure}

For example, in the undirected graph of Fig.\ \ref{fig:001}(a), the pairwise Markov property implies that $i\ci k\cd \{j,l\}$ and the global Markov property implies that $\{i,l\}\ci k \cd j$. 

We define a \emph{random network} to be the random graph $G=(V,E)$, where the node set $V$ consists of labeled individuals and the random tie set $E$, also called the set of \emph{dyads}, consists of binary random variables taking values in $\{ 0, 1\}$.  In a realization of a random network, nodes $i$ and $j$ are connected if the random variable corresponding to the tie $ij$ takes the value $1$, and they are disconnected otherwise.

\begin{figure}[h!]
            \centering
            \begin{tabular}{cc}
            \scalebox{0.15}{\includegraphics{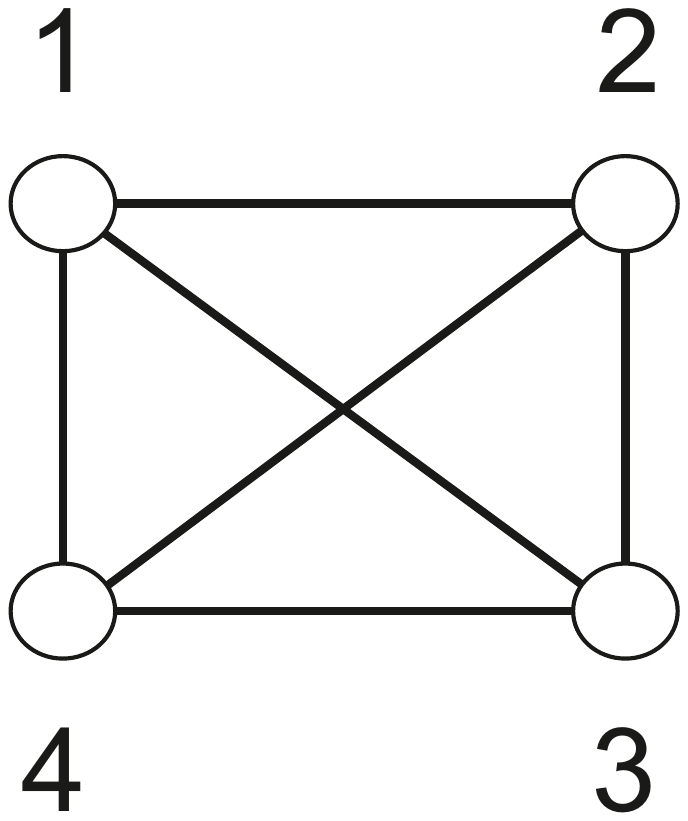}} &
            \scalebox{0.15}{\includegraphics{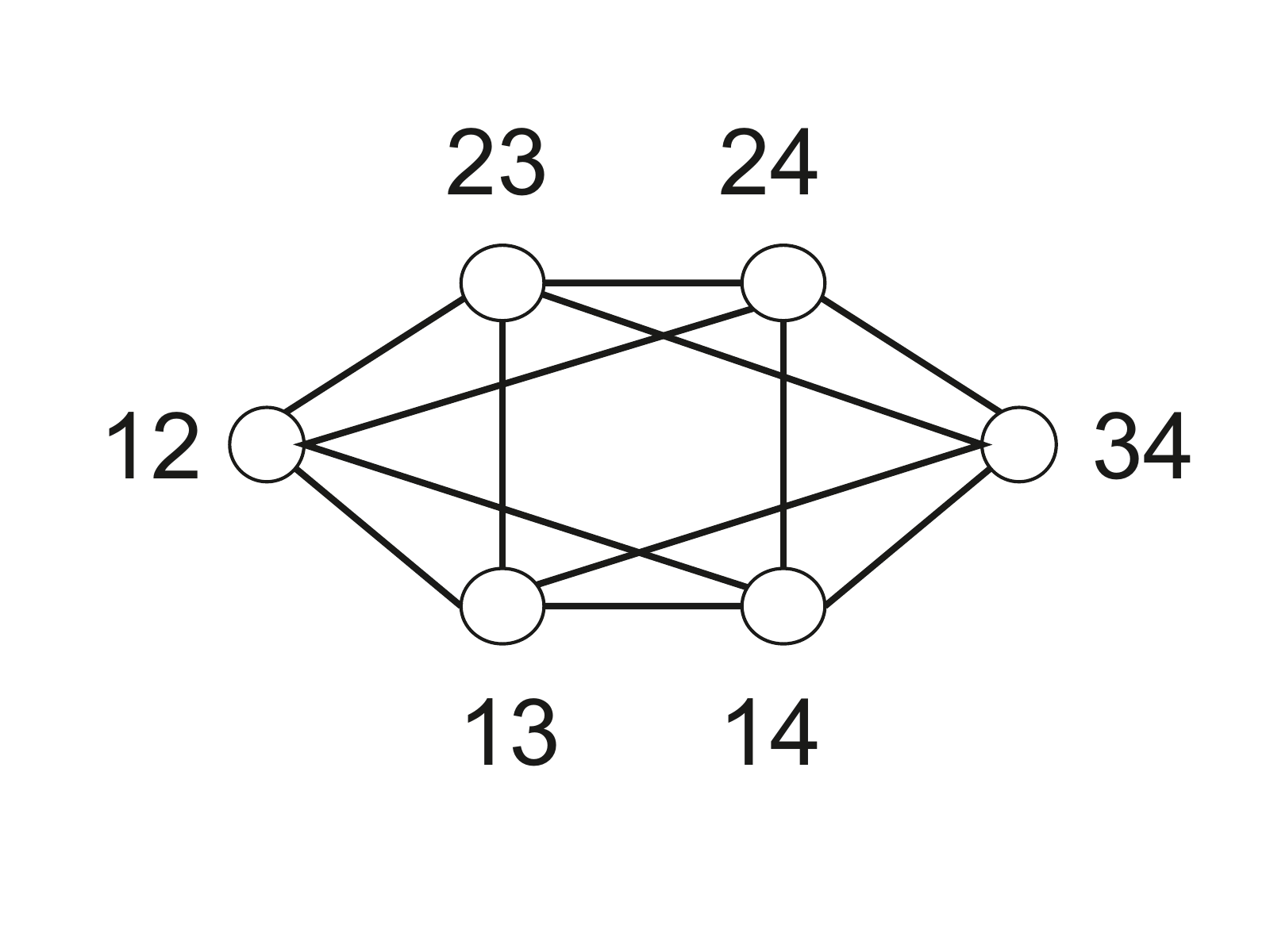}}\\
           $G$  & $D$
            \end{tabular}
            \caption{A random network $G$ with $4$ nodes and its corresponding line graph $D$.}
        \label{fig:3ex2}
\end{figure}

Now suppose that we are interested in networks  with $n=|V|$ nodes. Then, for a random network $G$ (which is a \emph{complete} random network in the sense that all dyads are existent) on $n$ nodes, it holds that its number of dyads is $|E|={n \choose 2}:=m$. Conditional independencies among the dyads of $G$ can then be expressed through a dependency graph $D$  on $m$ vertices, each vertex corresponding to a dyad in $G$.
We will only be concerned with certain types of conditional independencies, which we introduce next.  Let $i$, $j$, $k$, and $l$ be labels for the nodes of the random network $G$. We say that a dependency graph satisfies the \emph{Markov dependence property} if when ties $ij$ and $kl$ do not share a common node in $G$ then $ij \not\sim kl$ in $D$, i.e.\ vertices $ij$ and $kl$ are not adjacent. Notice also that the definition requires only non-neighboring ties of $G$ to be non-adjacent vertices in $D$ and not vice versa; therefore, the dependency graphs we propose could be any subgraph of the \emph{line graph} of $G$: the line graph $L(G)$ of a graph $G$ is the intersection graph of
the tie set $E$, i.e.\ its vertex set is $E$ and $e_1\sim e_2$ if and only if
$e_1$ and $e_2$ have a common endpoint \citep[p.\ 168]{wes01}. For example, for networks with $4$ nodes, all dependency graphs that satisfy the Markov dependence property are the subgraphs of the dependency graph $D$ depicted in Fig.\ \ref{fig:3ex2}. In addition to begin amenable to theoretical analysis, there are  practical justifications for adopting  such restrictions; see the discussion in Section \ref{sec:4}.

The type of restrictions on the dependency graphs described above is directly inspired by the Markov properties for networks put forward by \citet{fra86} in their seminal paper. Our modeling choice is, however, different in the two following ways: 1) In \citet{fra86}, a unique dependency graph (namely \emph{the line graph} of the complete graph  is used to model networks.  Here, on the other hand, we model any possible subgraph of the graph used in Frank and Strauss. Therefore, we deal with different possible independence structures that might occur for networks. 2) In \citet{fra86}, they assume exchangeability among the dyads of the network in order to reduce the number of parameters, whereas here we combine the graphical model with the known network models in the literature to obtain fewer parameters. This also ensures that our models inherit the desired properties of the baseline network model.

As we shall see in the next section, we are particularly interested in the \emph{cliques}, i.e.\ complete subgraphs, of the dependency graph. A \emph{triangle} in network $G$ is a subgraph consisting of nodes $i,j,k$ and ties $ij,jk,ik$. A \emph{$r$-star} in $G$ is a subgraph consisting of nodes $i,i_1,\dots,i_r$ and ties $ii_1,\dots,ii_r$;  we call the node $i$ the \emph{hub} of the star $C$, and write $i=\hub(C)$.

The following observation plays an important role in the paper: under Markov dependence property, cliques in $D$ correspond to stars and triangles in the random network $G$: A clique of size $1$, i.e.\ a vertex in $D$, is of form $ij$, and hence corresponds to the tie $ij$ in $G$, i.e.\ a $1$-star; a clique of size $2$, i.e.\ an edge in $D$, is of from $ij,ik$, and hence, because of the Markov dependence property, corresponds to the $2$-star with hub $i$ and other nodes $j,k$ in $G$; a clique of size $3$ that is of form $ij,ik,il$ in $D$ corresponds to a $3$-star in $G$; and a clique of size $r$, $r\geq 4$, corresponds to an $r$-star in $G$. A clique of size $3$ that is of form $ij,ik,jk$ in $D$ corresponds to a triangle in $G$; we call such cliques of $D$ \emph{hubless}.

Henceforth in this paper, we assume that a dependency graph is given, and the independence structure for the network is determined by this corresponding dependency graph. 
\section{Network models based on undirected hierarchical models}
\subsection{Undirected graphical models for networks}

Henceforth, let $\mathcal{G}_n$ be the set of all possible realizations of a network on $n$ nodes. We do not distinguish between the observed network $x \in \mathcal{G}_n$ itself and the binary vector of its dyads $x=(x_{12},\dots,x_{1n},x_{23},\dots,x_{2n},\ldots,x_{n-1,n}) \in \{0,1\}^{m}$, where every dyad is between two labeled nodes with labels $i$ and $j$ and $m = {n \choose 2}$.

Our goal is to model $P(x)$, the probability of observing $x$, given an undirected dependency graph $D$ on the set of $m$ dyads conforming to the Markov dependence property defined in the previous section. For this purpose, we will  use  \emph{hierarchical log-linear models}, which have been comprehensively studied for modeling undirected graphs in the graphical models literature; see \citet{lau96,bis07}.
In detail, let $\mathcal{C}$ 
be the set of all cliques in $D$, and let $\mathcal{C}_0=\mathcal{C}\cup\{\varnothing\}$. A hierarchical log-linear model corresponding to $\mathcal{C}_0$ can be written as
\begin{equation}\label{model1}
\log(P(x))=\sum_{C\in \mathcal{C}_0}u_C(x),\nn x\in\{0,1\}^m,
\end{equation}
where each $u_C(x)=u_{C}(x_C)$ is an appropriate function of $x \in \{0,1\}^m$ that depends only on the coordinates in $C$. In particular, $u_\emptyset$ is a constant function ensuring that $\sum_{x \in \{ 0,1\}^m} P(x) =1$.
 The hierarchical assumption requires that, if a term
$u_A(x)$ is set to zero for all $x$, so are all the terms $u_B(x)$ such that $A\subseteq B$. Hence,
the maximal cliques correspond to the maximal interaction terms not set to zero.
These are also called the \emph{generators} of the model.


We should first warn the reader not to confuse these models with hierarchical exponential random graph models, proposed by \citet{sch12}. Here, as will be seen in this section, the goal is to use hierarchical log-linear models to model networks with dependencies among the dyads. 

Since dyads are binary variables, the model \eqref{model1} can be parametrized as follows. Set, for each $C \in \mathcal{C}_0$ and $x \in \{0,1\}^m$
\begin{equation}\label{eq:u}
u_C(x)=\gamma_C \prod_{c\in C}x_c,
\end{equation}
were $\gamma_C \in \mathbb{R}$. Thus, for every clique, there exists only one parameter, $\gamma_C=u_C(1_C)$.There are $2^m-1$ equations (for every $x \in \mathcal{G}_n$ subject to the probabilities adding up to $1$) and $|\mathcal{C}|$ parameters in the model. It is easy to show that among all dependency graphs with $m$ vertices, the complete graph yields the largest number of parameters with $2^m-1$ parameters. 
Combining \eqref{model1} and \eqref{eq:u} yields the log-linear representation
\begin{equation}\label{eq:05}
P(x)=\tilde{u}\exp\{\sum_{C\in \mathcal{C}}\gamma_C\prod_{c\in C}x_c\}, \nn x\in \mathcal{G}_n,
\end{equation}
where $\tilde{u}=u_{\varnothing}$ is the normalizing constant, ensuring that the probabilities add up to $1$.

The representation (\ref{eq:05}) holds, of course, for any arbitrary binary graphical model. In the present setting however, where each point $x$ correspond to a network realization and the dependency graph satisfies the Markov dependency property defined above, the model \eqref{eq:05} can be interpreted using network statistics: Recall that  under Markov dependence property, cliques in $D$ correspond to stars and triangles in the random network $G$. Indeed, in the representation (\ref{eq:05}), for each clique $C$, the term $\prod_{c\in C}x_c$ is non-zero if and only if the subgraph of $G$ corresponding to the dyads $\{ c \in C\}$ is either a triangle of a star.

\begin{figure}[h!]
\centering
\includegraphics[scale=0.15]{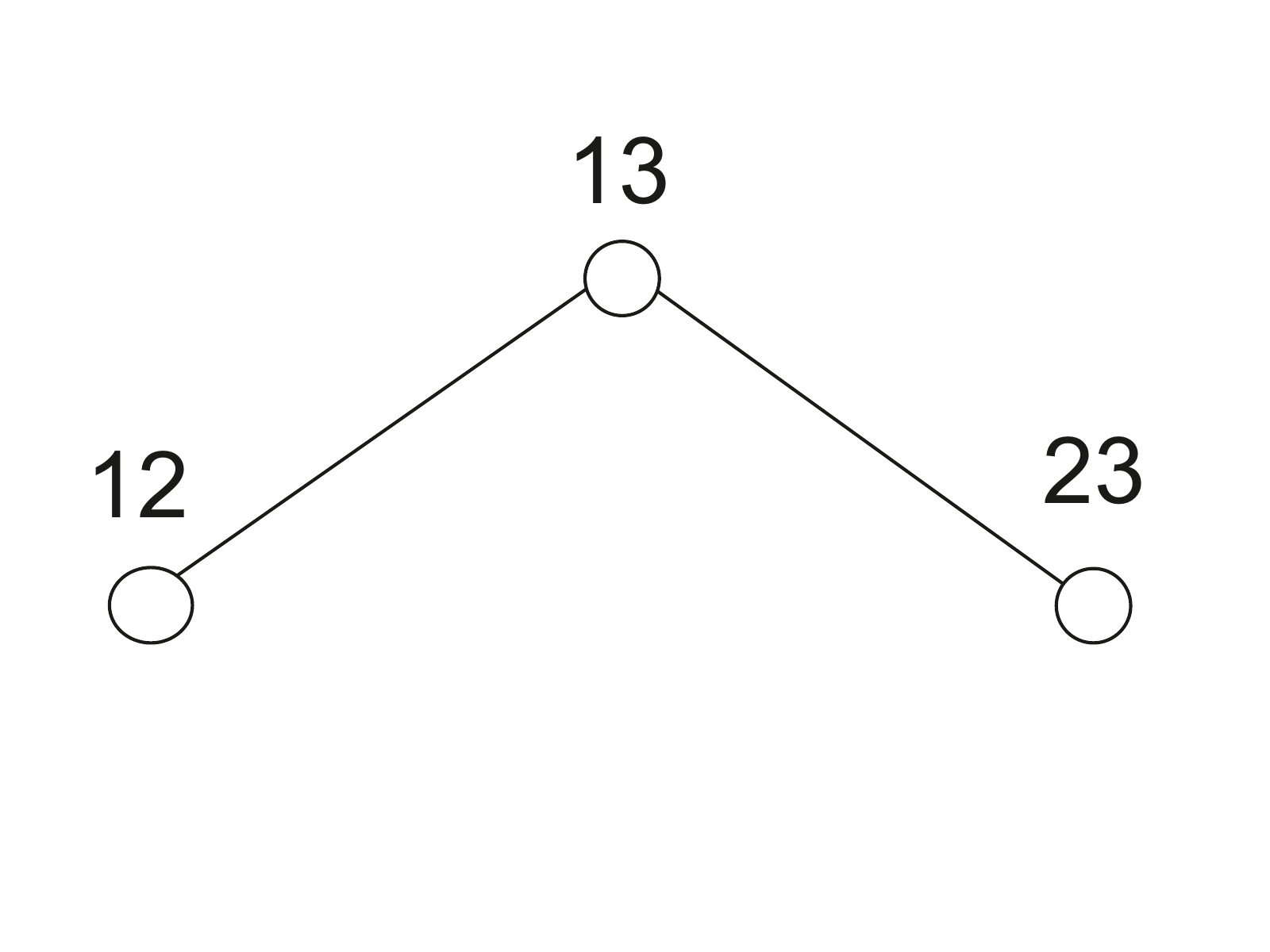}
\caption{A dependency graph corresponding to a network with node set $\{1,2,3\}$.}\label{fig:1}
\end{figure}

For example for the dependency graph in Fig.\ \ref{fig:1}, we have that
\begin{equation*}
P(x)=\tilde{u}\tilde{u}_{12}(x)\tilde{u}_{13}(x)\tilde{u}_{23}(x)\tilde{u}_{12,13}(x)\tilde{u}_{13,23}(x)=\tilde{u}\exp\{\gamma_{12}x_{12}+\gamma_{13}x_{13}+\gamma_{23}x_{23}+\gamma_{12,13}x_{12}x_{13}+\gamma_{13,23}x_{13}x_{23}\}.
\end{equation*}

In the next sections, we use the above model to generalize the
Erd\"{o}s-R\'{e}nyi and $\beta$-models in order to deal with the independence structure implied by the dependency graphs. This is done by putting constraints on the parameters $\gamma_C$ in (\ref{eq:05}) that come from the mentioned network models. The method is independent of the choice of the network models and can be applied to other network models that assume dyadic independence.
\subsection{Hierarchical Erd\"{o}s-R\'{e}nyi models}
In Erd\"{o}s-R\'{e}nyi models, it is assumed that ties occur independently, and the probabilities $p_{ij}$ of observing a tie between nodes $i$ and $j$ are all equal to $p$.  Hence the probability of observing a network is
\begin{equation*}
P(x)=\prod_{i<j}p^{x_{ij}}(1-p)^{1-x_{ij}}=(1-p)^{{n \choose 2}}\prod_{i<j}\exp\{x_{ij}\log(\frac{p}{1-p})\},\nn x\in \mathcal{G}_n.
\end{equation*}
In order to come up with an Erd\"{o}s-R\'{e}nyi type model that captures the independence structure implied by a given $D$, we generalize this model in the sense that the model for the baseline, where the dependency graph is the null graph, i.e.\ a graph with no edges, is the same as the model above. This implies that, in this case, one can consider $\tilde{u}_{ij}=e^{q x_{ij}}$, where $q=\log(p/(1-p))$, and leave $(1-p)^{{n \choose 2}}$ in the normalizing constant $\tilde{u}$.

In order to define the hierarchical Erd\"{o}s-R\'{e}nyi model, we set the following constraints on parameters in the model in (\ref{eq:05}) that conform with the constraints in Erd\"{o}s-R\'{e}nyi model. The remaining parameters after setting the constraints are denoted by the vector of parameters $q$ and a single parameter $t$.
\begin{equation}\label{eq:06}
\gamma_C=\left\{
                               \begin{array}{l}
                                 q^{(r)},\nn \text{if}\nn  C=\{ii_1,ii_2,\dots,ii_r\};\\
                                 t,\nn \text{if}\nn C=\{ij,ik,jk\}.
                               \end{array}
                             \right.
\end{equation}
An interpretation of the parameters is provided below. The hierarchical feature of the model is such that $q^{(r)}=0$ implies that $q^{(r+1)}=0$. In the saturated model, which corresponds to the line graph of the random network $K_n$, the number of parameters is $n$, and for dependency graphs with maximal clique of size $d<n$, the number of parameters is $d+1$.

In the example of Fig.\ \ref{fig:1}, under this model, we have
\begin{equation*}
P(x)=\tilde{u}\exp\{q^{(1)}x_{12}+q^{(1)}x_{13}+q^{(1)}x_{23}+q^{(2)}x_{12}x_{13}+q^{(2)}x_{13}x_{23}\}.
\end{equation*}
This, for example, implies that $12\ci 23\cd 13$, as is also implied by the dependency graph, since
\begin{equation*}
P(x)=\tilde{u}\exp\{q^{(1)}(x_{12}+x_{13})+q^{(2)}x_{12}x_{13}\}\exp\{q^{(1)}x_{23}+q^{(2)}x_{13}x_{23}\}.
\end{equation*}
Let $d$ be the size of the largest clique in $D$. By using (\ref{eq:06}), (\ref{eq:05}) can be written in exponential family form:
\begin{equation}\label{eq:07}
P(x)=\exp\{\sum_{r=1}^dq^{(r)}s^{(r)}_{\mathcal{C}^{(r)}}(x)+ts'_{\tau}(x)-\psi(q,t)\}, \nn x\in \mathcal{G}_n,
\end{equation}
where $\mathcal{C}^{(r)}$ is the set of all cliques with $r$ vertices in $D$, $q^{(r)}\in \mathbb{R}$, and $s_{\mathcal{C}^{(r)}}^{(r)}(x)$ is the number of $r$-stars in $x$ whose edges form a member of $\mathcal{C}^{(r)}$; similarly, $\tau$ is the set of all cliques with $3$ vertices in $D$ of form $\{ij,jk,ik\}$, and $s'_{\tau}(x)$ is the number of triangles in $x$  whose edges form a member of $\tau$.

Notice that $s^{(1)}_{\mathcal{C}^{(1)}}(x)$ is simply the number of ties of $x$ since all vertices of $D$ are considered cliques.

Therefore, since parameter $q^{(r)}$ corresponds to higher order interactions (of dimension $r$) in the dependency graph, it can be interpreted as propensity for the network to possess specific $r$-stars related to the cliques of the dependency graph. Hence, $q^{(1)}$ can be interpreted in the same way as the parameter $q$ in Erd\"{o}s-R\'{e}nyi. Similarly, $t$ can be interpreted as propensity for the network to possess triangles related to the dependency graph. Indeed, the sufficient statistics are correlated with each other, and the value $0$ for a parameter, say $q^{(r)}$,  means that given the value of other parameters, the number of cliques of size $r$ is close to the average number of possible cliques of size $r$.

Notice also that for the saturated model, we have that
\begin{equation*}
P(x)=\exp\{\sum_{r=1}^{n-1}q^{(r)}s^{(r)}(x)+ts'(x)-\psi(q,t)\},\nn x\in \mathcal{G}_n,
\end{equation*}
where $s^{(r)}(x)$ is the number of $r$-stars; and $s'(x)$ is the number of triangles.

For example, the model corresponding to the graph in Fig.\ \ref{fig:1}, can be written in exponential family form as
\begin{equation*}
P(x)=\exp\{q^{(1)}e(x)+q^{(2)}s_{\{\{12,13\},\{13,23\}\}}^{(2)}(x)-\psi(q)\}.
\end{equation*}

Obtaining the normalizing constant  in a closed form, for models in exponential family, in principle allows us to apply optimization methods for obtaining the maximum likelihood estimator. Notice that except in very few cases (such as Erd\"{o}s-R\'{e}ny and $\beta$-models), the normalizing constant in ERGMs is typically not in closed from.

Here we sum over all possible values of the binary vector $x$ in (\ref{eq:05}) after inserting the parameters in (\ref{eq:06}), and set it equal to $1$ in order to calculate the normalizing constant $\psi(q,t)$:
\begin{equation}\label{eq:081}
\psi(q,t)=\log(1+\sum_{r=1}^{n(n-1)/2}\sum_{H\in \mathcal{D}^{(r)}}\exp\{\sum_{r'=1}^{\min(d,r)}c^{(r')}(H)q^{(r')}+ c'(H)t\}),
\end{equation}
where $\mathcal{D}^{(r)}$ is the set of all subgraphs of $D$ with $r$ vertices, and $c^{(r')}(H)$ is the number of cliques of size $r'$ in $H$; and similarly $c'(H)$ is the number of cliques of size $3$ in $H$ of form $(ij,jk,ik)$. This could be written as $\psi(q)=\log((1+e^{q^{(1)}})^{{n \choose 2}}+f(q))$, where the term $\log(1+e^{q^{(1)}})^{{n \choose 2}}$ corresponds to cliques of size $1$, which are the vertices of the dependency graph. By neglecting the other term in the logarithm, we obtain the normalizing constant for the Erd\"{o}s-R\'{e}nyi model.

For example, for the dependency graph in Fig.\ \ref{fig:1}, we obtain
\begin{equation*}
\psi(q)=\log(1+3e^q+e^{2q}+2e^{2q+q^{(2)}}+e^{3q+2q^{(2)}}).
\end{equation*}

\subsection{Hierarchical $\beta$-models}
Next, we apply an approach  analogous to the one described in the previous section to
the $\beta$-model. The directed version of
$\beta$-model is the $p_1$-model \cite{hol81}, and the following approach can further generalize for $p_1$-model with few minor additional technicalities. For brevity, we have not included this in this paper.

In $\beta$-models, it is also assumed that ties occur independently, and the probability $p_{ij}$ of observing a tie between nodes $i$ and $j$ is parameterized as follows:
\begin{equation*}
p_{i,j}=\frac{e^{\beta_i+\beta_j}}{1+e^{\beta_i+\beta_j}},\nn \forall i\neq j,\nn \beta_1,\dots,\beta_n\in \mathbb{R}^n,
\end{equation*}
where $\beta_i$ can be interpreted as the propensity of node $i$ to have ties. The probability of observing a network is
\begin{equation*}
P_{\beta}(x)=\prod_{i<j}p_{ij}^{x_{ij}}(1-p_{ij})^{1-x_{ij}}=\prod_{i<j}e^{(\beta_i+\beta_j)x_{ij}}\frac{1}{1+e^{(\beta_i+\beta_j)}},\nn x\in \mathcal{G}_n.
\end{equation*}
In this case, the model above, which is the model for the baseline, can be considered to be $\tilde{u}_{ij}=e^{(\beta_i+\beta_j)x_{ij}}$ and $1/(1+e^{(\beta_i+\beta_j)})$ can be left in the normalizing constant $\tilde{u}$.

Now again suppose that there is a dependency graph $D$ that satisfies the Markov dependence property, and  is modeled by the hierarchical model (\ref{eq:05}).

We have observed in Section 2 that cliques in $D$ correspond to stars and triangles in $G$.
In order to define the hierarchical $\beta$-model, we set the following
constraints on the parameters in the model in (\ref{eq:05}) that conform with
the constraints in the $\beta$-model. The remaining parameters, after setting the constraints, are denoted by vectors of parameters $(\beta)_i$ and a single vector of parameter $\eta_i$.
\begin{equation}\label{eq:6}
\gamma_C=\left\{
                               \begin{array}{l}
                                 \beta_i^{(1)}+\beta_j^{(1)},\nn \text{if}\nn  C=\{ij\} \\
                                 \beta_i^{(r)},\nn \text{if}\nn  C=\{ii_1,ii_2,\dots,ii_r\},\nn r\geq 2 \\
                                 \eta_i+\eta_j+\eta_k,\nn \text{if}\nn C=\{ij,ik,jk\}.
                               \end{array}
                             \right.
\end{equation}

An interpretation of the parameters is provided below. As before, $\beta_i^{(r)}$ are hierarchical in the sense that if $\beta_i^{(r)}=0$ then $\beta_i^{(r+1)}=0$. In the saturated model, the number of parameters is $n^2$, but when the maximal clique size in the dependency graph is of size $d$, the number of parameters is $n\cdot d$.

In the example of Fig.\ \ref{fig:1}, under this model, we have
\begin{equation}\label{eq:14}
P(x)=\tilde{u}\exp\{(\beta_1^{(1)}+\beta_2^{(1)})x_{12}+(\beta_1^{(1)}+\beta_3^{(1)})x_{13}+(\beta_2^{(1)}+\beta_3^{(1)})x_{23}+\beta_1^{(2)}x_{12}x_{13}+\beta_3^{(2)}x_{13}x_{23}\}.
\end{equation}
This, for example, implies that $12\ci 23\cd 13$, as is also implied by the dependency graph, since
\begin{equation*}
P(x)=\tilde{u}\exp\{(\beta_1^{(1)}+\beta_2^{(1)})x_{12}+(\beta_1^{(1)}+\beta_3^{(1)})x_{13}+\beta_1^{(2)}x_{12}x_{13}\}\exp\{(\beta_2^{(1)}+\beta_3^{(1)})x_{23}+\beta_3^{(2)}x_{13}x_{23}\}.
\end{equation*}
By using (\ref{eq:6}), (\ref{eq:05}) can be written in exponential family form:
\begin{equation*}
P(x)=\exp\{\sum_{i=1}^{n}\sum_{r=1}^n\beta_i^{(r)}d_{i,\mathcal{C}_i^{(r)}}^{(r)}(x)+\eta_id_{i,\tau_i}(x)-\psi(\beta,\eta)\}, \nn x\in \mathcal{G}_n,
\end{equation*}
where $\mathcal{C}_i^{(r)}$ is the set of all cliques with $r$ vertices in $D$ such that all their vertices share $i$, and $d_{i,\mathcal{C}_i^{(r)}}^{(r)}(x)$ is the number of $r$-stars in $x$ with hub $i$ such that its endpoints pairing with $i$ form a member of $\mathcal{C}_i^{(r)}$; similarly, $\tau_i$ is the set of all cliques with $3$ vertices in $D$ of form $\{ij,jk,ik\}$, and $d_{i,\tau_i}(x)$ is the number of triangles in $x$ that contain $i$ and two other vertices such that they form a member of $\tau_i$.

Notice that $d_{i,\mathcal{C}_i^{(1)}}^{(1)}(x)$ is simply the degree of node $i$ since all vertices of $D$ are considered cliques.

Therefore, the parameter $\beta_i^{(r)}$ can be interpreted as the propensity of
node $i$ in the network to be the hub of specific $r$-stars related to the
cliques of the dependency graph. Hence, $\beta_i^{(1)}$ can be interpreted in
the same way as the parameter $\beta_i$ in the $\beta$-model. Similarly, $\eta_i$ can be interpreted as propensity for node $i$ in the network to possess triangles related to dependency graph.

For the saturated model, the model is
\begin{equation*}
P(x)=\exp\{\sum_{i=1}^{n}\sum_{r=1}^n\beta_i^{(r)}d_{i}^{(r)}(x)+\eta_id'_{i}(x)-\psi(\beta,\eta)\}, \nn x\in \mathcal{G}_n,
\end{equation*}
where $d_{i}^{(r)}(x)$ is the number of $r$-stars with $i$ as the hub; and $d'_{i}(x)$ is the number of triangles that contain $i$. In this case the sufficient statistics $d_{i}^{(r)}(x)$ are determined for $r\geq 2$ by $d_{i}^{(1)}(x)$. However, sufficient statistics in the submodels of the saturated model, $d_{i,\mathcal{C}_i^{(r)}}^{(r)}(x)$, can be arbitrary. For dense dependency graphs, the correlation between sufficient statistics can be high, and in some cases there might even be linear dependencies. Verifying this requires a case by case verification, generally, but for sparser dependency graphs this is not an issue.

For example, the model corresponding to the graph in Fig.\ \ref{fig:1}, can be written in exponential family form as
\begin{equation*}
P(x)=\exp\{\beta_1^{(1)}d_1^{(1)}(x)+\beta_2^{(1)}d_2^{(1)}(x)+\beta_3^{(1)}d_3^{(1)}(x)+\beta_1^{(2)}d_{1,\{\{12,13\}\}}^{(2)}(x)+\beta_3^{(2)}d_{3,\{\{13,23\}\}}^{(2)}(x)-\psi(\beta)\}.
\end{equation*}
As in the hierarchical Erd\"{o}s-R\'{e}nyi case, we are interested in writing  the normalizing constant  in a closed form in order to be able to apply optimization methods for obtaining the maximum likelihood estimator. We sum over all possible values of the binary vector $x$ in (\ref{eq:05}) after inserting the parameters in (\ref{eq:6}), and set it equal to $1$ in order to calculate the normalizing constant $\psi(\beta,\eta)$.
\begin{equation}\label{eq:81}
\psi(\beta,\eta)=\log(\prod_{i<j}(1+e^{\beta_i+\beta_j})+\sum_{r=2}^{n(n-1)/2}\sum_{H\in \mathcal{D}^{(r)}}(e^{\sum_{d=2}^{|V(H)|}\sum_{C\in\mathcal{C}^{(d)}(H)}\beta_{\hub(C)}^{(d)}+\sum_{C\in\tau(H)}(\eta_{c_1}+\eta_{c_2}+\eta_{c_3})}-1)e^{\sum_{v\in H}\beta_{v_1}+\beta_{v_2}}),
\end{equation}
where $\mathcal{D}^{(r)}$ is the set of all subgraphs of $D$ with $r$ vertices,
$V(H)$ is the vertex set of $H$, $\mathcal{C}^{(d)}(H)$ is the set of all cliques
in $H$ with $d$ vertices except the hubless cliques, $\tau(H)$ is the
set of all hubless cliques, which are of form $\{(i,j),(i,k),(j,k)\}$, and we
denote such $i$, $j$, and $k$ by $c_1$, $c_2$, and $c_3$. In addition, we write
$v$ as $(v_1,v_2)$. In (\ref{eq:81}), the term
$\prod_{i<j}(1+e^{\beta_i+\beta_j})$ corresponds to cliques of size $1$, which
are the vertices of the dependency graph, and neglecting the other term in the
logarithm, we obtain the normalizing constant for $\beta$-model, $\sum_{i<j}\log(1+e^{\beta_i+\beta_j})$.

Equation (\ref{eq:81}) can be used to compute the normalizing constant explicitly, although depening on the size of the network and the density of $D$, the computation of the sum could become intractable. For example, for the dependency graph in Fig.\ \ref{fig:1}, and from (\ref{eq:14}) we obtain
\begin{equation*}
\begin{split}
\psi(\beta)=\log(\sum_x\exp\{(\beta_1^{(1)}+\beta_2^{(1)})x_{12}+(\beta_1^{(1)}+\beta_3^{(1)})x_{13}+(\beta_2^{(1)}+\beta_3^{(1)})x_{23}+\beta_1^{(2)}x_{12}x_{13}+\beta_3^{(2)}x_{13}x_{23}\})\\
=\log(1+e^{\beta_1^{(1)}+\beta_2^{(1)}}+e^{\beta_1^{(1)}+\beta_3^{(1)}}+e^{\beta_2^{(1)}+\beta_3^{(1)}}+e^{\beta_1^{(1)}+\beta_2^{(1)}+\beta_1^{(1)}+\beta_3^{(1)}+\beta_1^{(2)}}+e^{\beta_1^{(1)}+\beta_2^{(1)}+\beta_2^{(1)}+\beta_3^{(1)}}\\ +e^{\beta_1^{(1)}+\beta_3^{(1)}+\beta_2^{(1)}+\beta_3^{(1)}+\beta_3^{(2)}}+e^{\beta_1^{(1)}+\beta_2^{(1)}+\beta_1^{(1)}+\beta_3^{(1)}+\beta_2^{(1)}+\beta_3^{(1)}+\beta_1^{(2)}+\beta_3^{(2)}})\\
=\log((1+e^{\beta_1^{(1)}+\beta_2^{(1)}})(1+e^{\beta_1^{(1)}+\beta_3^{(1)}})(1+e^{\beta_2^{(1)}+\beta_2^{(1)}})
+(e^{\beta_1^{(2)}}-1)e^{2\beta_1^{(1)}+\beta_2^{(1)}+\beta_3^{(1)}}\\+(e^{\beta_3^{(2)}}-1)e^{2\beta_3^{(1)}+\beta_1^{(1)}+\beta_2^{(1)}}+(e^{\beta_1^{(2)}+\beta_3^{(2)}}-1)e^{2\beta_1^{(1)}+2\beta_2^{(1)}+2\beta_3^{(1)}}).
\end{split}
\end{equation*}
The method proposed above for Erd\"{o}s-R\'{e}nyi and beta is not restricted to these models. In general, if there is a model that assumes dyadic independence then this method can be applied in the following manner:

Every linear exponential random graph model can be written in the form expressed in (\ref{eq:05}). The idea is that, for every clique $C$, the parameter $\gamma_C$ is further parametrized based on the baseline network model such that for every clique size $r$ there is a family of parameters $\{\theta_{i,k}\}_k\in\mathcal{K}$, where $\mathcal{K}$ is the set of parameters in the baseline network model. The baseline model corresponds to the empty dependency graph, and $\gamma_{ij}$, for every $i,j$, is reparametrized by the first order parameters $\{\theta_{1,k}\}_k\in\mathcal{K}$ in order to obtain the baseline network model.

In general, by using the reparametrization of $\gamma_C$, (\ref{eq:05}) can be written in exponential family form and the sufficient statistics will show up in the model, In addition, for these models,  if the normalizing constant of the baseline network model is in closed form then, by summing over all
possible values of the binary vector $x$ in the reparametrized version of (\ref{eq:05}), the normalizing constant can be written in closed form, although it still contains a sum over subgraphs of the dependency graph as opposed to a sum over all networks with $n$ nodes.

\section{Parameter estimation and simulation studies}\label{sec:3}
\subsection{Maximum likelihood estimation}
One important difference between the proposed models in this paper and other
exponential random graph models that do not assume dyadic independence is that the normalizing constants
(see (\ref{eq:081}) and (\ref{eq:81})) in our proposed models are in
\emph{closed form} in the sense that they do not depend on summation over all
networks. The models inherit this property from the models on which they are
based (i.e.\ the Erd\"{o}s-R\'{e}nyi and $\beta$-models). However, the normalizing constants depend on summation over subgraphs of the corresponding dependency graph, which can still be computationally demanding. We will, however, show below that for sparse dependency graphs some computations are manageable. This is essential for implementing the ML estimation for these models. This is in contrast to other exponential random graph models without dyadic independence assumption, which usually require some type of Markov chain Monte Carlo methods in order to compute the normalizing constant; see for example \citet{hun08}.

Henceforth, for brevity, we assume that there is no triangle of form $\langle ij,ik,jk\rangle$ in the dependency graph $D$. This implies that the parameter $t$ can be removed from the model. However, one can trivially generalize all the calculations and algorithms below to the case that $t$ is existent in the model.

By (\ref{eq:07}), the log-likelihood of the Erd\"{o}s-R\'{e}nyi model can be written as
\begin{equation*}
l(q)=\sum_{r=1}^{n-1}q^{(r)}s^{(r)}(x)-\psi(q),
\end{equation*}
where $\psi(q)$ is defined in (\ref{eq:081}).

This function is obviously concave. The goal is to apply the \emph{gradient descent} method (for example, see \citet{sny05}) to find its maximum. An element of the  gradient of the log-likelihood, for $m=n(n-1)/2$, is
\begin{equation}\label{eq:11}
 \partial l(q)/\partial q^{(i)}=s^{(i)}(x)-\frac{\sum_{r=1}^m\sum_{H\subseteq D^{(r)}}c^{(i)}(H)\exp\{\sum_{r'=1}^{\min(d,r)}c^{(r')}(H)q^{(r')}\}}{1+\sum_{r=1}^m\sum_{H\subseteq D^{(r)}}\exp\{\sum_{r'=1}^{\min(d,r)}c^{(r')}(H)q^{(r')}\}},
\end{equation}
where $d$ is the maximal clique size in $D$.
The main problem in computing the gradient vector is the large sum in both numerator and denominator, whose number of terms in the worst case is of order $O(n^3d2^m)$.
Here we will address this issue.

By calculating the term $r=1$ separately in (\ref{eq:11}), we obtain for $2\leq i\leq d$,
\begin{equation}\label{eq:20}
 \partial l(q)/\partial q^{(i)}=s^{(i)}(x)-\frac{\sum_{r=i}^m e^{rq^{(1)}}\sum_{H\subseteq D^{(r)}}c^{(i)}(H)\exp\{\sum_{r'=2}^{\min(d,r)}c^{(r')}(H)q^{(r')}\}}{1+me^{q^{(1)}}+\sum_{r=2}^m e^{rq^{(1)}}\sum_{H\subseteq D^{(r)}}\exp\{\sum_{r'=2}^{\min(d,r)}c^{(r')}(H)q^{(r')}\}};
\end{equation}
and
\begin{equation}\label{eq:21}
 \partial l(q)/\partial q^{(1)}=s^{(1)}(x)-\frac{me^{q^{(1)}}+\sum_{r=2}^m re^{rq^{(1)}}\sum_{H\subseteq D^{(r)}}\exp\{\sum_{r'=2}^{\min(d,r)}c^{(r')}(H)q^{(r')}\}}{1+me^{q^{(1)}}+\sum_{r=2}^m e^{rq^{(1)}}\sum_{H\subseteq D^{(r)}}\exp\{\sum_{r'=2}^{\min(d,r)}c^{(r')}(H)q^{(r')}\}}.
\end{equation}

Suppose that $D$ is \emph{sparse} in the sense that the number of non-isolated vertices in $D$, denoted by $m'$, is of order of a constant, i.e.\ not growing with $n$, or it is increasing by a rate slower than $n$. This implies that, based on our assumption, there is only a finite number of individuals in the model to which the connected ties are dependent. By increasing $n$, the dependency graph converges to the empty graph, and the model converges to  Erd\"{o}s-R\'{e}nyi. Denote the induced subgraph by non-isolated vertices by $D'$. An expression in the last term in the denominator of (\ref{eq:20}) or (\ref{eq:21}) is expanded as follows:
\begin{multline*}
\sum_{H\subseteq D^{(r)}}\exp\{\sum_{r'=2}^{\min(d,r)}c^{(r')}(H)q^{(r')}\}=\sum_{r''=2}^{\min(m',r)}{m-m'\choose r-r''}\sum_{H\subseteq D'^{(r'')}}\exp\{\sum_{r'=2}^{\min(d,r'')}c^{(r')}(H)q^{(r')}\}+{m-m'\choose r}+m'{m-m'\choose r-1},
\end{multline*}
where ${a \choose b}:=0$, for $b>a$.

Therefore, by the same argument for numerators, (\ref{eq:20}) and (\ref{eq:21}) can  be written as
{\small
\begin{multline}\label{eq:22}
 \partial l(q)/\partial q^{(i)}= s^{(i)}(x)-\frac{\sum_{r''=i}^{m'}\sum_{H\subseteq D'^{(r'')}}c^{(i)}(H)\exp\{\sum_{r'=2}^{\min(d,r'')}c^{(r')}(H)q^{(r')}\}(\sum_{r=2}^m{m-m'\choose r-r''}e^{rq^{(1)}})}{1+\sum_{r=1}^me^{rq^{(1)}}({m-m'\choose r}+m'{m-m'\choose r-1})+\sum_{r''=2}^{m'}\sum_{H\subseteq D'^{(r'')}}\exp\{\sum_{r'=2}^{\min(d,r'')}c^{(r')}(H)q^{(r')}\}(\sum_{r=2}^m{m-m'\choose r-r''}e^{rq^{(1)}})};
\end{multline}}
and
{\small
\begin{multline}\label{eq:23}
 \partial l(q)/\partial q^{(1)}= s^{(1)}(x)-\frac{\sum_{r=1}^mre^{rq^{(1)}}({m-m'\choose r}+m'{m-m'\choose r-1})+\sum_{r''=2}^{m'}\sum_{H\subseteq D'^{(r'')}}\exp\{\sum_{r'=2}^{\min(d,r'')}c^{(r')}(H)q^{(r')}\}(\sum_{r=2}^m r{m-m'\choose r-r''}e^{rq^{(1)}})}{1+\sum_{r=1}^me^{rq^{(1)}}({m-m'\choose r}+m'{m-m'\choose r-1})+\sum_{r''=2}^{m'}\sum_{H\subseteq D'^{(r'')}}\exp\{\sum_{r'=2}^{\min(d,r'')}c^{(r')}(H)q^{(r')}\}(\sum_{r=2}^m{m-m'\choose r-r''}e^{rq^{(1)}})},
\end{multline}}
where ${a \choose b}:=0$, for $b<0$.

By fixing $q^{(1)}$, and assuming $m'$ is of order of a constant, since $d$ can no longer grow with $n$, the number of terms in the sum in both numerator and denominator is of order $O(C2^n)$, where $C$ is a constant. Now for every value of $q^{(1)}_*$, by the gradient descent method, the largest value for $g(q^{(1)}_*)=\max_{q^{(2)},\dots,q^{(d)}} l(q^{(1)}_*,q^{(2)},\dots,q^{(d)})$ can be computed. (If $f(x,y)$ is concave then so is $f(x,y_*)$ for a fixed value of $y_*$.) Hence, one can use iterative methods in optimization to find the value of $q^{(1)}$ within a certain range with a certain precision that maximizes the likelihood function.


\subsection{Computational and simulation study}
We have written code that calculates the maximum likelihood estimator by the optimization method discussed above, although, at the moment it cannot computationally carry large networks ($>50$ nodes) (the computational difficulty  depends also on the density of the dependency graph).

We have also incorporated the corresponding statistics in the package \verb"ERGM" \citep{hun07}, by coding up the change statistics in the package \verb"ERGM.userterterms" \citep{hun13}, and can now utilize the functionality provided in the \verb"ERGM" package such as those for approximating the maximum likelihood estimator by Markov chain Monte Carlo methods and simulating networks based on exponential random graph models.

For a given network $x$ and dependency graph $D$, and in order to have some base for comparisons regardless of the difference in the  density of the simulated networks, we find the  \emph{likelihood ratio} under the maximum likelihood estimator of hierarchical Erd\"{o}s-R\'{e}nyi and Erd\"{o}s-R\'{e}nyi (the latter can be exactly computed easily):
\begin{equation}\label{eq:LR.stat}
S=2l_{HER}(\hat{q}_1,\dots,\hat{q}_r;x,D)-2l_{ER}(\hat{q};x).
\end{equation}
Since the Erd\"{o}s-R\'{e}nyi is a submodel of hierarchical Erd\"{o}s-R\'{e}nyi, it holds that $l_{HER}>l_{ER}$, and hence $S>0$. In addition, for two fixed hierarchical models, denoted by $1$ and $2$, and the same network, the difference between the log-likelihoods is proportional to $S_1-S_2$.
%

We first provide a method to simulate a vector of size $m$ of binary variables with conditional correlations (linear dependencies) induced by a given undirected dependency graph $D$. By this method the dependencies induced by the dependency graph are preserved, but there might be independencies that turn into non-linear dependencies among the simulated binary data:

We simulate an $m\times m$ symmetric positive-definite matrix $M$ (e.g., using of The QR Decomposition of a Matrix in \verb"LAPACK" \citep{and99}). We then change the values of the matrix in order  for zeros to correspond to missing edges of $D$ as follows.
For $A \ci B \cd C$, it holds that $\cov(A,B)=\cov(A,C)\cov(C,C)^{-1} \cov(C,B)$. Thus, start with $M$ as a covariance matrix and cycle through
all pairs corresponding to the missing edges of $D$ by forcing the formula to hold for $(i,j,V\setminus\{i,j\})$. We stop when the sum of the deviations from the concentrations fitted to the missing edges (i.e.\ the sum of the corresponding elements on the generated matrix) is smaller than some default. For more details, see \citet{spe86}. Call the resulting matrix $L$.

We can now consider $K=L^{-1}$ to be the \emph{concentration matrix}, i.e.\ the inverse of the covariance matrix, of a Gaussian distribution with $0$ mean that is Markov to $D$.
We generate binary random variables by thresholding a normal distribution, i.e.\ by setting the simulated value of a variable to $0$ if the corresponding value in the Gaussian distribution is negative, and to $1$ if the corresponding value in the Gaussian distribution is positive. The linear conditional dependencies are preserved under thresholding; see \citet{lei98} and the R package \verb"bindata", introduced in its appendix.
%
%

We now assume that a simulated $x$ is an observed network with every element corresponding to a tie. We find an approximation of the maximum likelihood estimator for the hierarchical Erd\"{o}s-R\'{e}nyi model by the \verb"ERGM" package, and also the maximum likelihood estimator for Erd\"{o}s-R\'{e}nyi. We then calculate the likelihood ratio $S$. We discard the simulated networks for which the maximum likelihood estimator does not seem to exist (as implied by \verb"ERGM").

Here we conduct our study for several dependency graphs with complete connected components of different sizes (including isolated vertices); see also Section \ref{sec:4} for a short discussion on model selection (based on nodal attribute or exchangeability). We performed four simulations experiments.

\begin{figure}[h!]
            \centering
            \begin{tabular}{cc}
            \scalebox{0.25}{\includegraphics[height=25cm,width=27cm]{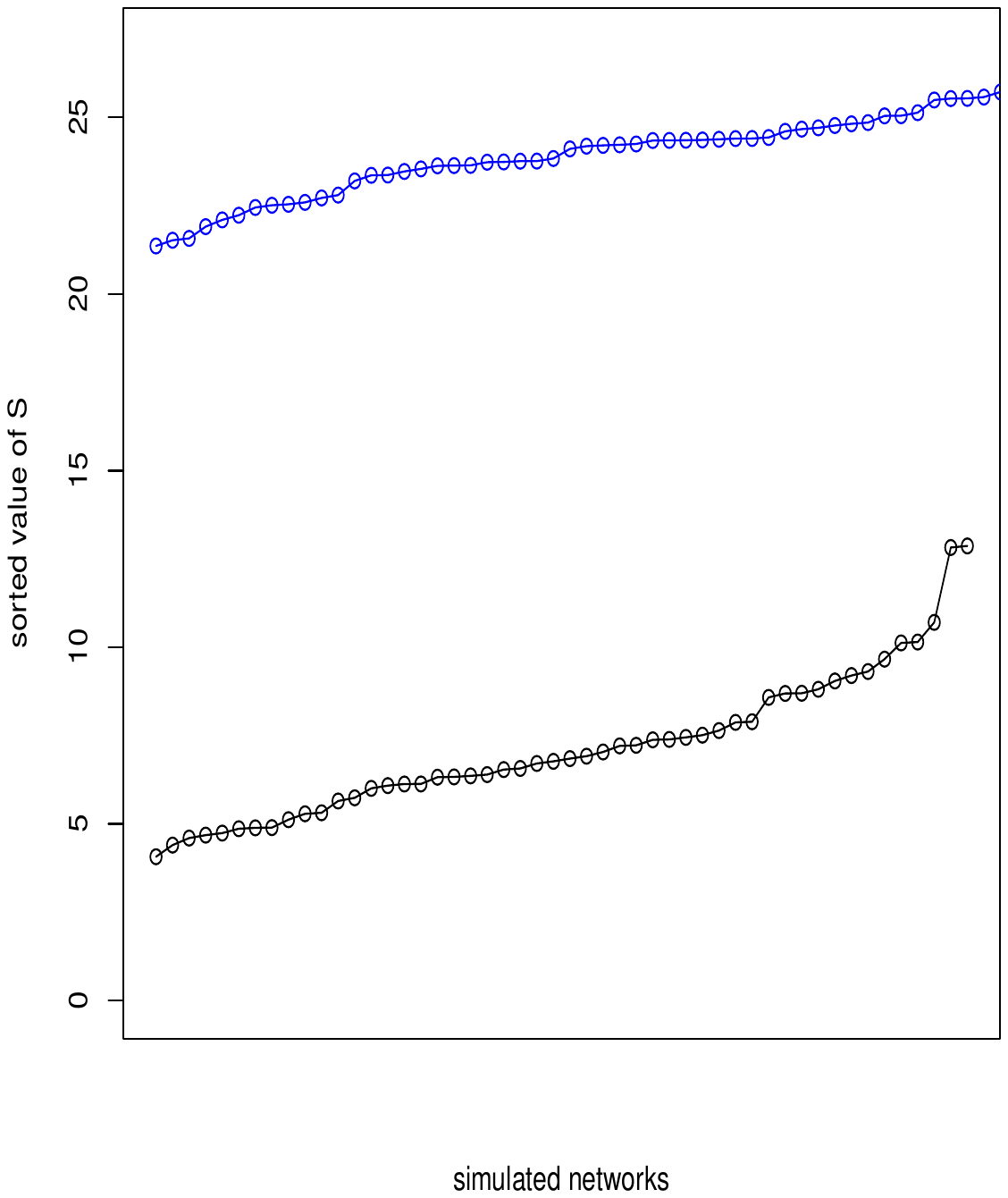}} &
            \scalebox{0.25}{\includegraphics[height=25cm,width=27cm]{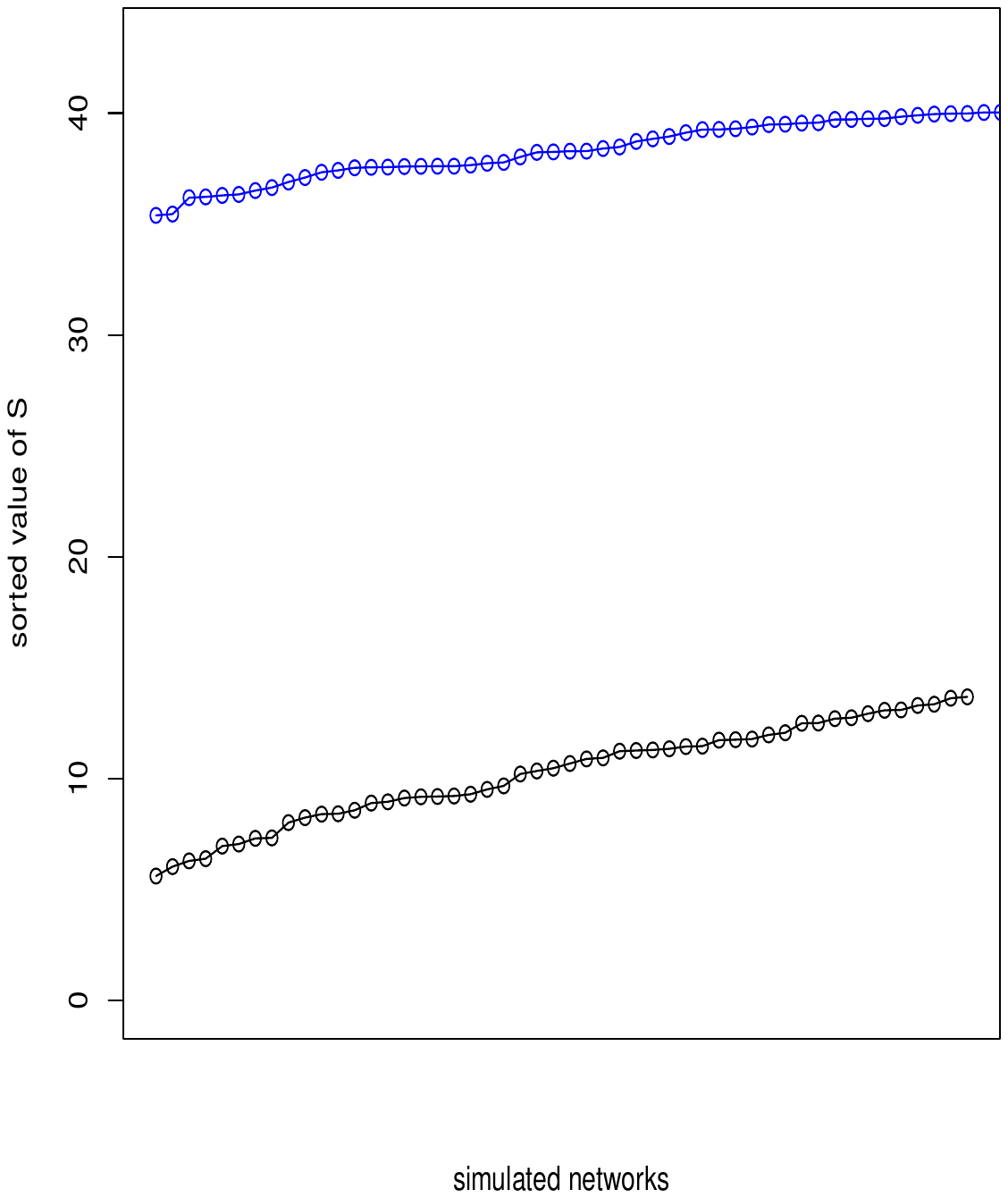}}\\
           (a)  & (b)
            \end{tabular}
            \caption{Ordered likelihood-ratios of networks simulated from a conditional independence structure implied by a dependency graph with $1225$ vertices and (a) with one clique of size $25$ together with isolated vertices, low correlations (black, at the bottom) and high correlations (blue, on top); (b) with one clique of size $40$ together with isolated vertices, low correlations (black, at the bottom) and high correlations (blue, on top).}
        \label{fig:12l}
\end{figure}

1) We simulate networks on $50$ nodes from the same dependency graph, but both using low values of the partial correlations $\rho_{ij,ik}$, for all possible vertices $ij$ and $ik$, among the dependent variables ($|\rho_{ij,ik}|<0.1$), and also  high values of the partial correlations ($|\rho_{ij,ik}|<0.9$). We consider two dependency graphs: one consisting of a clique of size $25$ and isolated vertices and the other of a clique of size $40$  and isolated vertices.
For each simulated network in each of the scenarios we compute the likelihood ratio statistic as in \eqref{eq:LR.stat}, and plot the ordered values of the corresponding statistics, in order to give a sense of the spread and tail of the their distribution. The results are presented in Figure \ref{fig:12l}. We clearly observe that the model performs better for higher correlations, which is expected.

\begin{figure}[h!]
            \centering
            \scalebox{0.25}{\includegraphics[height=25cm,width=27cm]{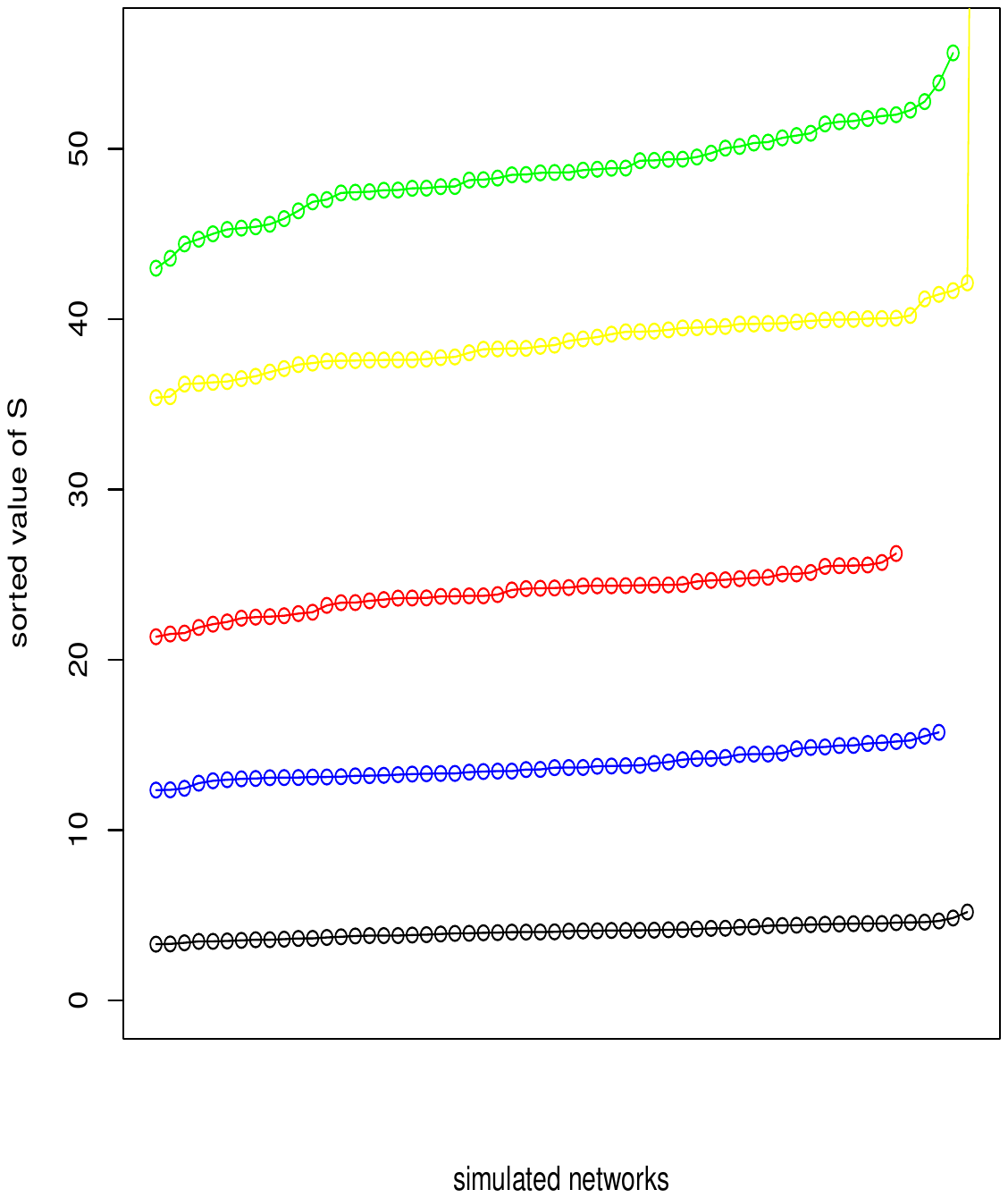}}
            \caption{Ordered likelihood-ratios of networks simulated from a conditional independence structure implied by a dependency graph with $1225$ vertices and one clique of size $5$ (black, at the bottom), $15$ (blue, second from the bottom), $25$ (red, in the middle), $40$ (yellow, second from the top), and $50$ (green, on top), and isolated vertices.}
        \label{fig:12}
\end{figure}

2) We simulate networks on $50$ nodes with different dependency graphs with medium level correlations. The dependency graph again consists of one cliques and isolated nodes, for increasing values  of the clique size (and therefore of the number of parameters in the model). It is obvious that the likelihood ratio should increase as well, which is confirmed by our study. Figure \ref{fig:12} shows the results of the experiment for cliques of size $5$, $15$, $25$, $40$, and $50$.

\begin{figure}[h!]
            \centering
            \begin{tabular}{cc}
            \scalebox{0.25}{\includegraphics[height=25cm,width=27cm]{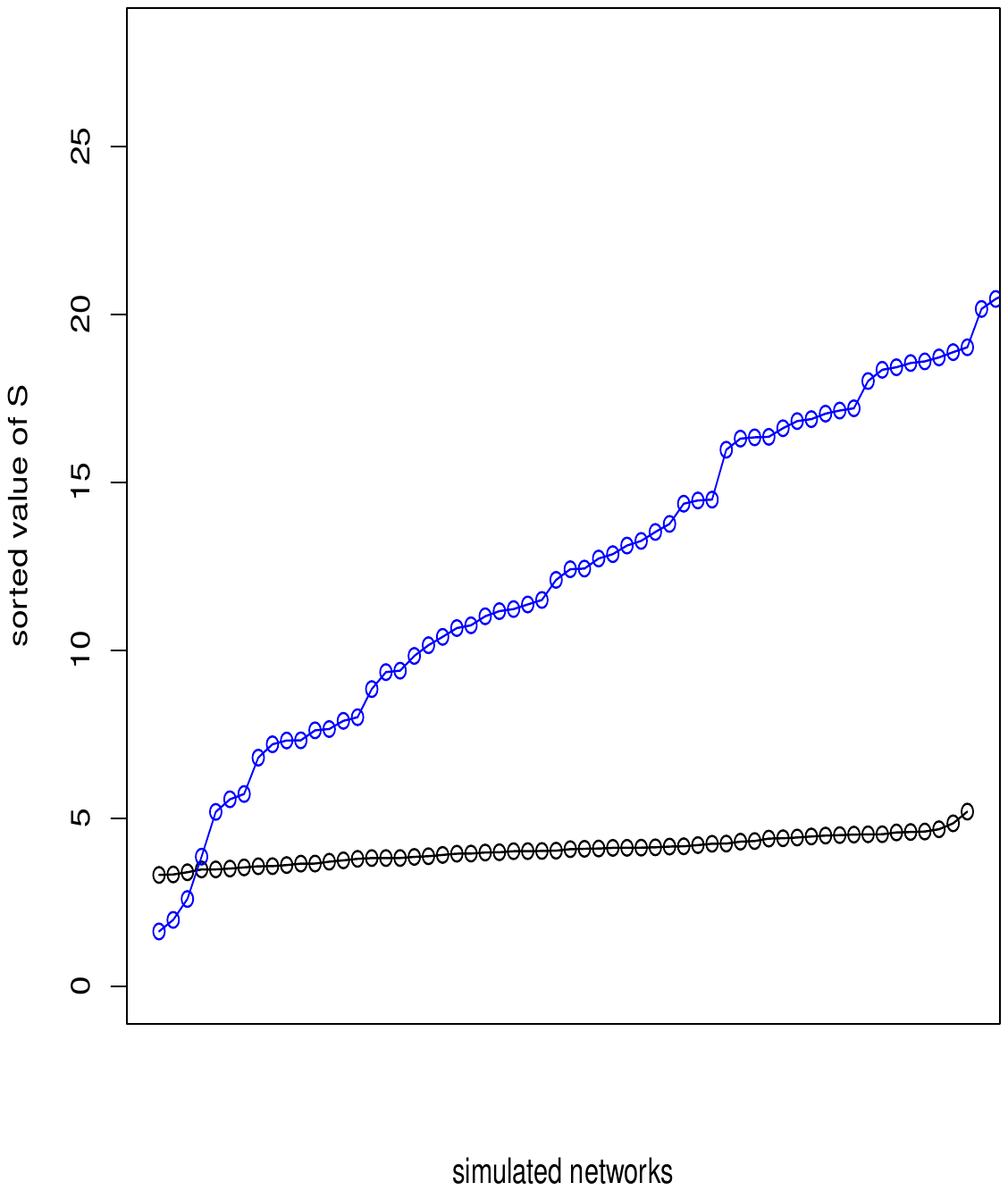}} &
            \scalebox{0.25}{\includegraphics[height=25cm,width=27cm]{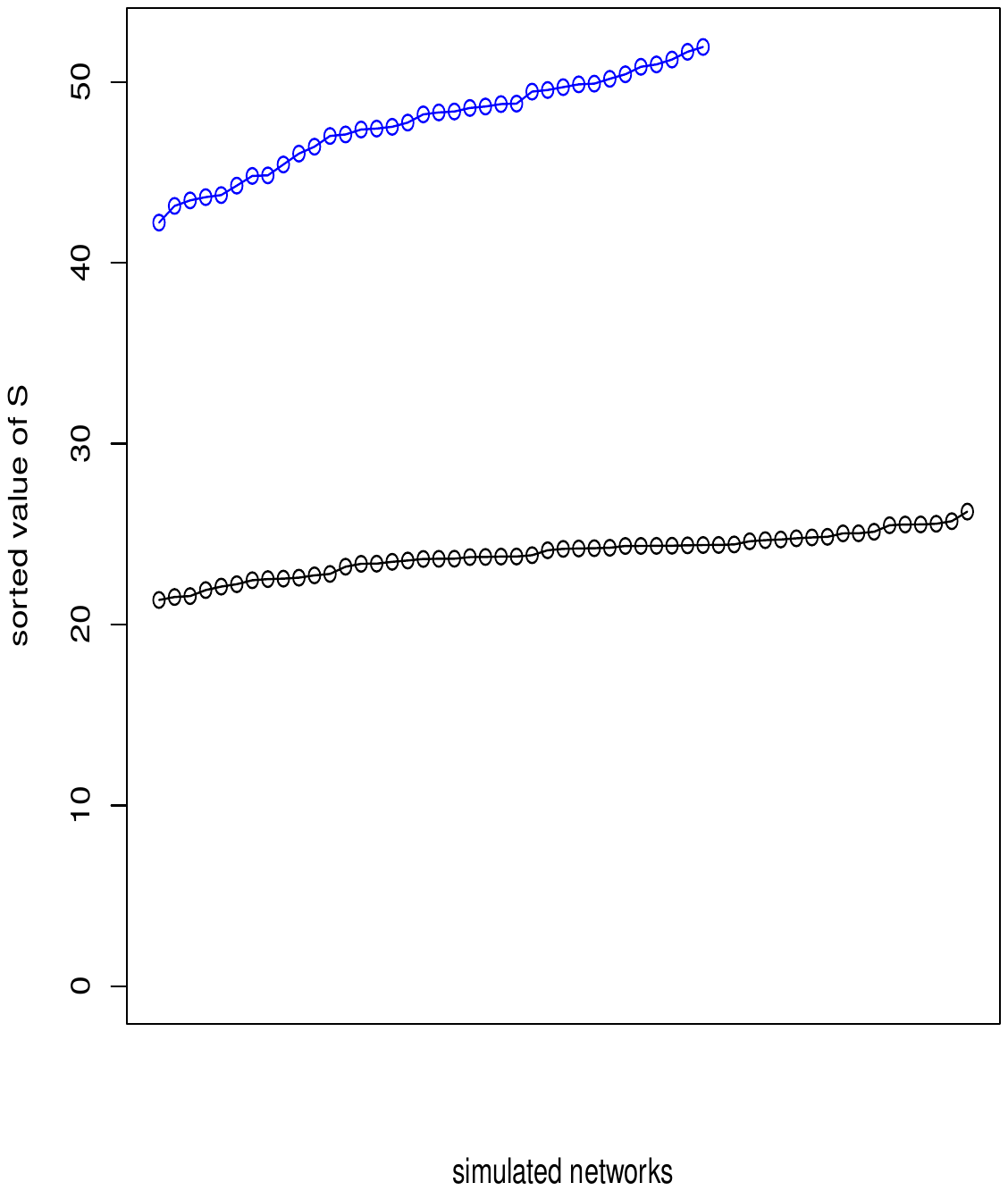}}\\
           (a)  & (b)
            \end{tabular}
            \caption{Ordered likelihood-ratios of networks simulated from a conditional independence structure implied by a dependency graph with $1225$ vertices and (a) one clique of size $10$ (black, at the bottom) and $5$ cliques of size $10$ (blue, on top) together with isolated vertices; (b) one clique of size $25$ (black, at the bottom) and $2$ cliques of size $25$ (blue, on top) together with isolated vertices.}
        \label{fig:3}
\end{figure}

3) The following study further suggests that the model works. We again simulate networks of the same size, but with different dependency graphs of the same maximal clique size. Hence, the number of parameters is the same for different dependency graphs. Here we compare models for two dependency graphs, one of which contains only one clique and isolated vertices, and the other has several disconnected cliques of the same size. In particular, we depict the values of the likelihood ratio for networks with $50$ nodes simulated from two dependency graphs that, along with isolated vertices, have respectively $1$ maximal clique of size $5$, and $10$ maximal cliques of size 5 (part (a) of Fig.\ \ref{fig:3}), and $1$ component of size $25$, and $2$ components of size 25 (part (b) of Fig.\ \ref{fig:3}). In this experiment we  discard cases where the maximum likelihood estimator does not exist (or at least cannot be computed by this method). It is seen  in  Fig.\ \ref{fig:3} (b) that the number of simulated networks, for which the maximum likelihood estimator seems to exist is visibly lower for the denser dependency graph. We observe the same trend in our simulation studies, where the simulated networks based on the models with denser dependency graphs are more affected by the issue of non-existence of the maximum likelihood estimator. This can be due to the fact that the value for certain  statistics in some simulated networks is $0$, or due to degeneracy issues in other cases.

As it is often the case with exponential random graph models, the issues of existence of the maximum likelihood estimator and
degeneracy \citep{Schweinberger2011InstabilitySA,aleejs,hand03} affect also the models we propose. In fact, in our simulations we have
encountered several cases in which the likelihood function does not appear to
be strongly
convex, as the estimated Fisher information matrix becomes nearly singular along optimizing sequences of parameters
with norms diverging to infinity. This is a clear indication of a
nonexistent maximum likelihood estimator and manifests itself in failed convergence of the Markov chain Monte Carlo maximum likelihood estimation procedure
and slow mixing. While the investigation of these important issues is clearly outside the scope of
this paper, nonetheless we recommend using small dependency graphs and a careful
monitoring of the convergence of the optimizing procedure of choice.


In addition, the model with more cliques seems to have a higher variance than the model with only one clique.

\begin{figure}[h!]
            \centering
            \scalebox{0.25}{\includegraphics[height=25cm,width=27cm]{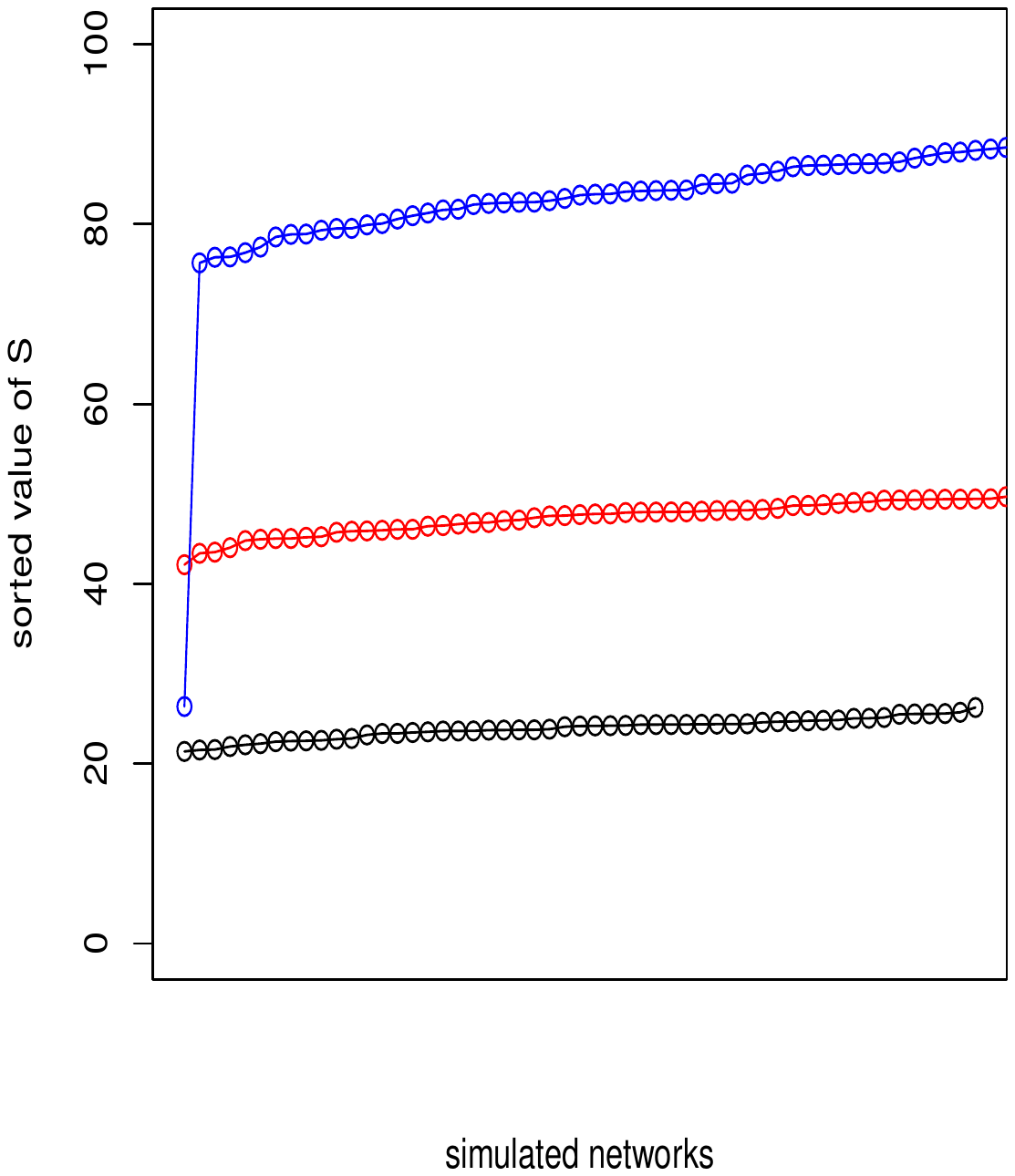}}
            \caption{Ordered likelihood-ratios of networks simulated from a conditional independence structure implied by a dependency graph with $m={n \choose 2}$ vertices and  one clique of size $n/2=25$ (black, at the bottom), $n/2=50$ (red, in the middle), and $n/2=100$ (blue, on top) together with isolated vertices.}
        \label{fig:120}
\end{figure}

4) We increase the number of nodes and try to keep the density of dependency graph (whose size increases of order of $n^2$) unchanged. As expected, we see that the value of $S$ increases when increasing the number of nodes. Here we depict examples of this value for simulated networks of size $50$, $100$, and $200$ in the graph of Fig.\ \ref{fig:120} with dependency graphs that contain a clique of size $n/2$ along with isolated vertices.
\section{Discussion and future work}\label{sec:4}
Our aim in this paper was to introduce a new approach for proposing a new family of models, where each model is the combination of a known network model and the hierarchical model for a given dependency graph. There are, however, still many open questions regarding the proposed models as well as natural ways to generalize such approach.

%
%
%
%
%
%
As discussed before, for these hierarchical models,  if the normalizing constant of the baseline network model is in closed form then the normalizing constant is in closed form, i.e.\ it does not contain a sum over all networks of size $n$. This is an essential aspect of these models, which makes exact parameter estimation possible. However, the normalizing constant still contains a sum over subgraphs of the dependency graph, which depending on the size and density of dependency graph could be computationally intractable.  As discussed in Section \ref{sec:3}, because of this computational demands, it is not possible to only apply standard optimization methods to find the maximum likelihood estimator. In this paper, we provided a fairly more sophisticated computational techniques that work for ``sparse" dependency graphs.

The question of how to efficiently find solutions for gradients (\ref{eq:22}) and (\ref{eq:23}) is indeed an interesting optimization problem. By skipping the $1$ in the denominator, a general form for the gradient that one should deal with is $\sum c_if_i/\sum f_i$. The other term in the gradient is computationally manageable.

Throughout this paper we worked under the assumption that there is a given dependency graph. In general, the model depends on the labeling of the network, i.e.\ the individuals represented by the network. In practical cases, a sparse dependency graph may normally be provided by experts, by marking certain individuals whose relations with other individuals affect the whole or some parts of the network.

If such an expert opinion is not available, there may be several methods to select a dependency graph based on the nature of the network or the model: It is common to observe some nodal attributes along with an observed network. In this case these attributes can be used for selecting a dependency graph.

In addition, features of networks could lead to specific dependence graphs. These include \emph{homophily}, i.e.\ individuals with similar characteristics being more
likely to relate \cite{kri09}, \emph{heterophily}, where individuals relate to those with different characteristics (as seen in some social networks \cite{loz14}), \emph{exchangeability}, in the sense that the model is invariant under relabeling of the nodes, and \emph{transitivity}, which states that given that we know whether $j$ and $k$ are friends (i.e.\ adjacent), knowing whether $i$ is a friend of $j$ would impact  the probability of $i$ being a friend of $k$.

For exchangeable ERGMs, it was shown in \cite{lau17} that the corresponding dependence graph could be either empty, complete,  the line graph of the complete graph, or its compliment.

For the case of transitivity, for three nodes, this corresponds to $ij\ci ik\cd jk$ not being satisfied, but the marginal independence $ij\ci ik$ being satisfied. This can be represented by the \emph{bidirected dependency graph} $ij\arc jk\arc ik$. This motivates introducing a parallel theory as in this paper to define a set of network models that preserve the independence structures of bidirected graphical models as opposed to undirected graphical models:  for a \emph{bidirected graph} $D$, where all edges are depicted as bidirected edges, $\arc$, the pairwise Markov property states  that for two non-adjacent  vertices $i$ and $j$, $i\ci j$, i.e.\ $i$ and $j$ are marginally independent; whereas the global Markov property states that $A\ci B\cd C$ when every path between disjoint vertex subsets $A$ and $B$ has no vertex in $C$.

Suppose that a given dependency graph captures the independence structure of a network of friendships. It is plausible to assume that the existence of a tie between individuals $i$ and $j$ and the existence of a tie between $k$ and $l$ are marginally independent, i.e., knowing whether $i$ and $j$ are friends would not change the probability of friendship between $k$ and $l$ when there is no information available on the existence of other friendships in the network. This conforms with the missing edge between $ij$ and $kl$ in a bidirected dependency graph that satisfies the Markov dependence property.  Moreover, a possible edge between $ij$ and $ik$ in the dependency graph indicates that knowing whether $i$ has a friend $j$ would impact the probability of $i$ having another friend $k$.

For the purpose of modeling ERGMs based on bidirected dependency graphs, 
\emph{marginal binary models} of \citet{drt08} can be used. The parameters of this model are marginal probabilities for every connected subgraph of the dependency graph (as opposed to every clique in hierarchical models). These models are  \emph{curved} exponential families. In order to make the parameter estimation more practical, one may use other marginal models introduced in graphical models that are in curved exponential family form; see, for example, the \emph{multivariate logistic transformation}, introduced in \citet{lup09}, and \emph{log-mean linear models}, introduced in \citet{rov13}. Of course, instead of the plain marginal binary models, the idea is to apply these models to networks.
\section*{Acknowledgements}\label{acknowledgements}
The authors are grateful to Sonja Petrovi\'{c}, Thomas Richardson, Despina Stasi, and Ryan Tibshirani for very helpful discussions.

%
%

\selectlanguage{english}
\bibliography{bib}

\end{document}